\documentclass[conference]{IEEEtran}
\IEEEoverridecommandlockouts
\usepackage{cite}
\usepackage{amsmath,amssymb,amsfonts}
\usepackage{algorithmic}
\usepackage{graphicx}
\usepackage{hyperref}
\usepackage{enumitem}
\usepackage{arydshln}
\usepackage{caption, multirow}
\usepackage{textcomp}
\usepackage{xcolor}
\def\BibTeX{{\rm B\kern-.05em{\sc i\kern-.025em b}\kern-.08em
    T\kern-.1667em\lower.7ex\hbox{E}\kern-.125emX}}
\begin{document}

\title{Data Augmentation for Drum Transcription with Convolutional Neural Networks}

\author{\IEEEauthorblockN{C\'eline Jacques}
\IEEEauthorblockA{\textit{Analysis-Synthesis Team} \\
\textit{STMS-UMR 9912, IRCAM, Sorbonne University, CNRS}\\
Paris, France \\
celine.jacques@ircam.fr}
\and
\IEEEauthorblockN{Axel R\"obel}
\IEEEauthorblockA{\textit{Analysis-Synthesis Team} \\
\textit{STMS-UMR 9912, IRCAM, Sorbonne University, CNRS}\\
Paris, France \\
axel.roebel@ircam.fr}

}

\maketitle

\begin{abstract}
A recurrent issue in deep learning is the scarcity of data, in particular precisely annotated data. Few publicly available databases are correctly annotated and generating correct labels is very time consuming. The present article investigates into data augmentation strategies for Neural Networks training, particularly for tasks related to drum transcription. These tasks need very precise annotations. This article investigates state-of-the-art sound transformation algorithms for remixing noise and sinusoidal parts, remixing attacks, transposing with and without time compensation and compares them to basic regularization methods such as using dropout and additive Gaussian noise. And it shows how a drum transcription algorithm based on CNN benefits from the proposed data augmentation strategy.
\end{abstract}


\section{Introduction}
\label{sec:intro}

The present paper deals with data augmentation strategies and their application to automatic transcription of drum events with deep neural networks given a limited set of training examples.

Automatic drum transcription in music recordings consists in annotating the time position at which a drum events occurred and specifying which drum instrument was played. 

In \cite{SCHLUETER2014}, they noticed that the onset detector based on Convolutional Neural Network (CNN) learned different features for harmonic onsets and percussive onsets. So the problem of drum transcription can be seen as a detection of qualified onsets and has therefore close relationship to the more general problem of finding onsets in music. For the onset detection task, deep learning reached the best results in 2017 at MIREX onset detection task over all the MIREX evaluation campaigns\footnote{\url{https://nema.lis.illinois.edu/nema_out/mirex2017/results/aod/}}. 

On the other hand, drum transcritpion can also be considered as a problem in itself. While some methods present very good results, e.g. \cite{PAULUS2009} with Hidden Markov Model (HMM), methods based on deep learning seem to be more and more common. In \cite{VOGL2016a}, a RNN is trained to calculate the activation functions of different drum instruments. The first study of CNN for drum transcription was performed in \cite{SOUTHALL2017}. Now CNN are widely used for drum detection \cite{Vogl2017, Jacques2018} and got the best results during the MIREX2018 drum transcription challenge\footnote{\url{https://www.music-ir.org/mirex/wiki/2018:Drum_Transcription_Results}}.

A recurrent issue for deep learning is the scarcity of annotated data. In image processing for example, if an object has always been presented to the network with the same orientation, it may encounter difficulties to detect it when the object is moved in other direction. For music similar issues appear when the available databases do not cover sufficient possibilities. The way an instrument sounds varies greatly from one music track to another. In fact, because getting precise annotation of data is very difficult and generating them time consuming, the amount of publicly available data with precise annotations will hardly contain as much variety as is present in the real world. For training Deep Neural Network (DNN) nearly all the information comes from annotated data. Therefore the more data with relevant variability one can provide, the better algorithm will work.

Besides other strategies such as student/teacher paradigm \cite{Wu2017}, data augmentation can be a solution to bridge the lack of data. It consists in applying different transformations on annotated sounds to create new data labeled with similar precision. However it is necessary that the augmented data keeps the features of the label. As all information is encoded in the data, high quality sound transformation is essential to avoid training the network on irrelevant artifacts that are introduced by means of the transformation algorithm. In extreme cases training on such artifacts may lead to algorithms which work only with transformed data. Because human perception is the ultimate target when playing music instruments, it appears a reasonable first approach to validate augmented data by means of listening. 

Some transformation methods with preservation of label meanings have been proposed in \cite{McFee2015, SCHLUETER2015} such as pitch shifting or time stretching. \cite{Vogl2018} investigates into  transcribing 18 classes of drum instruments. Because the database does not present a balanced  distribution of the percussive onsets of the different classes, it is re-synthesized to obtain a better distribution of the 18 classes. In \cite{Jacques2018}, training data are created by synthesizing sounds from midi files using three different soundfonts. In \cite{SCHLUETER2015} data augmentation is achieved by transforming the magnitude spectrogram of the input by scaling the spectrogram images. In \cite{Salamon2017}, the MUDA toolbox \cite{McFee2015} is used to transform environmental signals for the task of environmental sound classification. 

In this article, the use of 4 different transformations of sound signals will be evaluated in the context of  a drum transcription task. The transformations are particularly selected to transform features that seem relevant for detecting drum events: remixing attacks, remixing noise and sinusoidal parts, as well as transposition with and without time compensation and with varying degrees of transformation of the spectral envelop. It will be compared to two other methods often used for regularisation of DNN: the use of dropout and adding a white noise to the input spectrograms. 

In this article, we also focus on the choice of augmentation strategies and their parameters. The recall is always expected to increase if some variability is introduced in the database by data augmentation while precision might decrease if the parameter of data augmentation are not carefully chosen. 

The article is organized as follows: Section \ref{sec:method} presents the model on which the studied networks are based and the different transformations applied to the database. Section \ref{sec:drum_trans} presents results with thoses augmented databases. We finally draw conclusions in section \ref{sec:concl}.

\section{Methodology}\label{sec:method}

\subsection{Pre-processing}\label{subsec:preprocess}

The networks are fed with a Multi-Channel Mel Spectrograms (MCMS) representation \cite{Jacques2018} initially proposed in \cite{SCHLUETER2014}. MCMS represents the data in three log-magnitude mel band spectrograms. The short-time Fourier transform (STFT) is created with a hop size of 10ms and three window sizes: 23, 46 and 93ms. The spectrograms are then filtered with an 80 mel-band filter covering the bands from 27.5Hz to 16kHz. Thus the input data are three mel-band spectrograms with 80 mel-bands and different time resolutions. 

\subsection{CNN topology}\label{subsec:cnn_topo}

\begin{figure*}[ht]
    \center
    \includegraphics[scale=0.7]{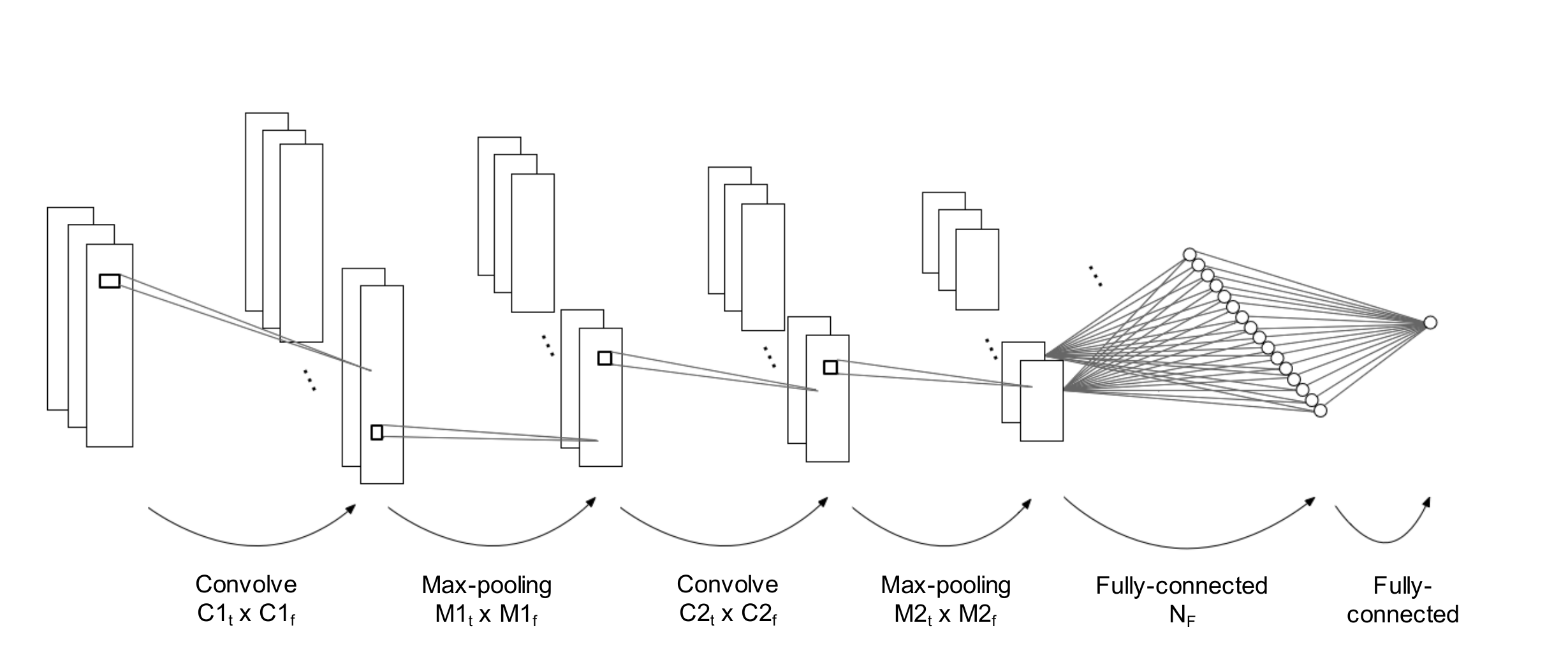}
    \caption{\it Network topology for onset detection and drum transcription.}
    \label{fig:cnn_detect}
\end{figure*}

The networks used in this article are all based on the CNN topology introduced in \cite{SCHLUETER2014}. The network topology shown in Fig. \ref{fig:cnn_detect} alternates two stacks of convolutional layers with ReLU activation followed by max-pooling layers and ends with a fully connected hidden layer with ReLU units and the output layer containing either a sigmoid unit or a linear unit. 

Each layer has different hyper-parameter such as filter sizes, number of filters, etc. The parameterization will be specified by means of a single character designating the type of layer operation (C: convolutional, M: max-pooling, F: fully-connected) followed by two or three numbers. In case of convolutional layers, the numbers specify filter dimension in time and frequency followed by the number of filters, for max-pooling, the pooling strides in time and frequency dimensions are specified, and for the fully connected layer, simply the number of hidden units. For the network in figure \ref{fig:cnn_detect}, the topology can be specified as
\begin{equation}
\begin{split}
&\text{C$_1$: }\text{C1}_{t}\times \text{C1}_{f}|\text{N}_{\text{C1}}, \text{M$_1$: }\text{M1}_t\times \text{M1}_f, \\
&\text{C$_2$: }\text{C2}_t\times \text{C2}_f|\text{N}_{\text{C2}}, \text{M$_2$: }\text{M2}_t\times \text{M2}_f, \\
&\text{F: }\text{N}_{\text{F}}
\end{split}\label{cnn:onset_baseline}
\end{equation}
That is a convolutional layer with $\text{N}_{\text{C1}}$ filters of size $\text{C1}_{t}\times \text{C1}_{f}$ followed by a max-pooling layer with strides $\text{M1}_t\times \text{M1}_f$, then a convolutional layer with $\text{N}_{\text{C2}}$ filters of size $\text{C2}_t\times \text{C2}_f$ followed by a max-pooling layer with strides $\text{M2}_t\times \text{M2}_f$ and in the last hidden layer $\text{F, }\text{N}_{\text{F}}$ nodes fully-connected to the previous feature maps.

\subsection{Basic regularization strategies}\label{sec:regularisation}

The strategies of data augmentation investigated in this article are compared to two regularization strategies: dropout and additive Gaussian noise. Because these methods are used similarly for image and sound processing, they are considered as data-independent methods in \cite{SCHLUETER2015}. 

The use of dropout during the training consists in setting some of outputs of a layer to zero with a given probability. Because the dense layers connects all the neurons together, it potentially creates extremely specialized hidden units which in turn may lead to an over-fitting of the training database. Randomly cancelling some hidden units increases redundancy and avoids over-specialization of  the network which is expected to learn more robust features. We note $P_d$ the probability of dropout.

An other way to introduce variability in data and to strengthen the network to variations is to add Gaussian noise to the input spectrogram. The Gaussian noise of mean zero is defined by its standard deviation which we note $\sigma$. 

Both data-independent method of regulation are applied on-the-fly during the training and do not intervene during evaluation and testing steps.

\subsection{Data augmentation for audio}\label{sec:data_augment}

To provide the networks with augmented data, we experiment four transformations: remixing the noise part, remixing attacks, transposing with or without time compensation including various degrees of transposition of the spectral envelope. All transformations are applied directly to the audio signal before pre-processing to convert it into the MCMS representation. 

The four transformations described below are all performed with the signal transformation kernel available in version 3 of the AudioSculpt program \cite{Bogaards2004} that can be scripted and controlled via the Unix command line: 
\begin{itemize}[noitemsep, topsep=1mm, leftmargin=3mm]
    \item \textbf{Remix noise}: After classifying sinusoidal and noise peaks in the signal audio \cite{Zivanovic2004}, they can be separated by means of spectral masking and remixed with selected mixing ratio. This leads to a modification of the balance between sinusoidal and noise components. Attacks, or onsets in general and percussive instruments in particular, contain different amounts of noise and transient sinusoids (resonances). Remixing them changes the energy distribution between these two components and can therefore be seen as a transformation of onset properties.
    Fast onsets will neither be detected as sinusoid nor as noise and will be covered by the following effect.\\
    \emph{Parameter:} $r_n$  designates the remixing factor applied to the spectral peaks detected as noise. 
    \item \textbf{Remix attacks}: After detecting the time-frequency locations of transient bins \cite{Roebel2003}, these bins are scaled by applying a linear factor to either soften or increase the transient strength. 100ms after each transient, the factor fades to one. This transformation aims to modulate the attack properties, notably in comparison to the subsequent stationary or release phase of the sound event.\\
    \emph{Parameter:} $r_a$  designates the remixing factor applied to the spectral peaks detected as transients. 
    \item \textbf{Transposition with and without time compensation}: Transposition is achieved by means of resampling the signal to a new sample rate without changing the sample rate used to play the sound. The anti-aliasing filter used has 3db bandwidth up to 90\% of the respective sample rate, and aliasing attenuation over -70dB.
    The resampling will scale all temporal relations in the signal leading notably to tempo changes, and faster evolution of attack transitions. These time scale changes can be compensated by applying a phase vocoder approximately following \cite{Laroche1999} additionally  making use of the transient preservation described in \cite{Roebel2003}.\\
    \emph{Parameter:} $t$ transposition in cents that is achieved by means of the re-sampling operation and $t\text{ } nc$ for transposition without time compensation.
    \item \textbf{Spectral envelope transposition}: For the transposition with and without time compensation, an additional transformation is added. The spectral envelope is estimated \cite{Roebel2005} and transposed while the pitch remains unchanged. This last effect leads to time varying filters and sound color changes. It has the potential to further increase data variability. \\
    \emph{Parameter:} $t_e$ transposition in cents of the spectral envelope. 
\end{itemize}


Data augmentation will therefore be performed with 5 different setups:
noise remixing only, attack remixing only, transposition with time compensation and envelope transposition, transposition without time compensation and envelope transposition and finally all the transformation combined. The parameterization of the resampling operation will be used equivalently for transposition with and without time compensation and the combination with envelope transposition will always be performed covering all possible combinations.

\subsection{Evaluation strategy} \label{sec:crossval}

The evaluation of the data augmentation strategies is performed independently for each type of transformation by means of cross database validation experiments \cite{Livshin2003}.  We train the CNN models on all but one databases and evaluate it on the remaining database. The training databases are augmented by one of the augmentation strategies.  

The selection of the result of a training cycle is selected by means of early stopping. For this a few tracks of the original training databases are selected as validation data  and all of the augmented sounds based on this validation subset are removed from the augmented versions.
Evaluation with these  examples serves to determine the best network from all the networks that have been obtained during the training phase. The validation set contains only original non-augmented data.

The experiments are realised with 3 different seeds for the initialization of the networks. The results are the average over those three initializations. 

\section{Application to drum transcription}\label{sec:drum_trans}

For automatic drum transcription, we compare the different transformations on the training data of MIREX 2018 drum transcription challenge with a cross-fold validation. As what is usually done in literature, we focus on the three main drum instruments: hi-hat (hh), bass-drum (bd) and snare-drum (sd). Those three instruments are responsible of most of the basic rhythms in western music. 

\subsection{CNN model}\label{subsec:cnn_model}

To perform the transcription of the three main drum instruments, three individual networks are trained independently. The three independent networks are based on the architecture of the network shown in fig. \ref{fig:cnn_detect}. Each network aims to detect one and only one drum instrument. They all have the same parameters given by the topology following eq. \ref{cnn:onset_baseline}. 
\begin{equation}
    \begin{split}
        &\text{C$_1$: }7\times 3|10, \text{ M$_1$: }1\times 3 \\
        &\text{C$_2$: }3\times 3|20, \text{ M$_2$: }1\times 3 \\
        &\text{F: }256 
    \end{split}
\end{equation}
To learn the weights, we use cross-entropy as loss function with sigmoid activation and Adam optimizer. 

\subsection{Augmented databases}

\textbf{Transformation parameters}

Each transformation is controlled by a parameter. This parameter is chosen so that the augmented version of an event is still recognizable as the same event. They are given in Table \ref{tab:param_mirex} and explained in sec. \ref{sec:data_augment}. For each value of the parameter $t$, the data are processed with all values of $t_e \in [-300, -200, -100, 100, 0, 100, 200, 300]$. 

The probability of dropout on the third layer and the standard deviation of the Gaussian noise added to the input are also given in Table \ref{tab:param_mirex}. 

In the following, the reference is the networks trained without any kind of data augmentation, i.e. no dropout, no additive Gaussian noise and trained only on original data. 

\begin{table}[!ht]
\centering
\begin{tabular}{c|c}
    Factor & Values \\\hline
    $P_d$ & 0.25, 0.5, 0.75 \\
    $\sigma$ & 0.05, 0.1, 0.2 \\
    \hdashline
    $r_n$ & 0.1, 0.3, 0.6, 1.5, 2, 3\\ 
    $r_a$ & 0.6, 1.5, 2, 3\\
    $t$   & -300,-200,-100,100,200,300\\
    $t\text{ } nc$ & -300,-200,-100,100,200,300
\end{tabular}\caption{\it Transformation parameters for MIREX 2018 database.}
\label{tab:param_mirex}
\end{table}

\textbf{MIREX2018: Drum transcription training database}

This dataset was provided for MIREX 2018 drum transcription challenge to train models. It is detailed on MIREX 2018 website\footnote{\url{https://www.music-ir.org/mirex/wiki/2018:Drum_Transcription\#Training_Data}}. It is made up of four subsets from different databases: 2005, GEN, MEDLEY and RBMA. They all contain annotated polyphonic pieces of music of different genres as well as drum only tracks.

From 3 hours of recordings, we obtain after transformation: about 19 hours for remix noise, about 12,5 hours for remix attacks and about 113 hours for transposition with time compensation and 113 hours without. 

\subsection{Experiments}

We give the results for each augmented version in the Table \ref{tab:drum_transcript}. The results are given as the mean of the Recall, Precision and F-measures over the four cross validation steps and over the three different initialization. We ran five experiments. 
\begin{enumerate}[noitemsep, leftmargin=5mm]
    \item \textbf{Original} We train the three CNNs detailed in section \ref{subsec:cnn_model} on the original data without dropout or additive Gaussian noise.
    \item \textbf{Dropout} We change the value of the probability of the dropout
    \item \textbf{Gaussian noise} Three values of standard deviation are tested to generate Gaussian noise which are added to the input spectrograms.
    \item \textbf{Audio transformation} The fourth experiment investigates the influence of the four data augmentation strategies explained at the section \ref{sec:data_augment}.
    \item \textbf{Transformations combined} The networks are trained on original data, the four augmented databases.
\end{enumerate}

The drums are considered as correctly detected if they are closer than 50ms from the ground-truth onsets.

\subsection{Results}

\begin{table}[!ht]
\centering
\begin{tabular}{ p{0.55cm}| p{0.45cm} | p{0.35cm} p{0.35cm} p{0.35cm}| p{0.35cm} p{0.35cm} p{0.35cm}| p{0.35cm} p{0.35cm} p{0.35cm} }
  \multicolumn{2}{c|}{}& \multicolumn{3}{c|}{\textbf{BD}} & \multicolumn{3}{c|}{\textbf{SD}} & \multicolumn{3}{c}{\textbf{HH}} \\
\multicolumn{2}{c|}{} & \textbf{R}  & \textbf{P} & \textbf{F} &\textbf{R}  & \textbf{P} & \textbf{F}& \textbf{R}  & \textbf{P} & \textbf{F} \\
\hline
  \multicolumn{2}{c|}{Orig.} & 87.4 & 73.8 & 80.2 & 67.5 & 54.0 & 59.0 & 78.6 & 64.3 & 69.8 \\
  \hline
   Drop.       & 0.25 & 88.2 & 74.2 & 80.6 & 66.5 & 54.3 & 58.9 & 78.6 & 64.0 & 69.6 \\
$P_d$          & 0.50 & 88.2 & 74.4 & 80.5 & 65.7 & 53.8 & 58.3 & 77.4 & 64.8 & 69.5 \\
              & 0.75 & 88.4 & 73.9 & 80.2 & 67.3 & 53.3 & 58.7 & 78.6 & 64.5 & 69.8 \\
\hline
Gaus.         & 0.05 & 87.8 & 73.7 & 79.9 & 67.3 & 54.1 & 59.2 & 79.8 & 63.8 & 70.0 \\
noise         & 0.1  & 87.8 & 74.3 & 80.3 & 67.0 & 52.8 & 58.3 & 79.9 & 63.5 & 70.0 \\
$\sigma$      & 0.2  & 88.1 & 73.8 & 80.1 & 67.2 & 52.7 & 58.2 & \textbf{79.8} & \textbf{64.1} & \textbf{70.2} \\
\hline
\multirow{5}{*}{Trans.} &rn    & 90.2 & 73.5 & 81.6 & \textbf{65.8} & \textbf{55.6} & \textbf{59.8} & 78.4 & 64.3 & 69.7 \\
&ra           & 90.2 & 75.0 & 81.7 & 67.0 & 53.7 & 59.0 & 78.7 & 63.9 & 69.5 \\
&t            & 91.2 & 70.0 & 80.7 & 69.4 & 53.5 & 59.5 & 80.4 & 61.1 & 68.4 \\
&t nc         & \textbf{90.0} & \textbf{75.7} & \textbf{82.1} & 68.3 & 53.6 & 59.4 & 80.4 & 61.7 & 68.8 \\
\hdashline
&All  & 90.6 & 74.0 & 81.2 & 67.7 & 54.3 & 59.6 & 79.6 & 62.3 & 68.9
\end{tabular}\caption{\it Results (F-measure) of drum transcription on MIREX 2018 database with dropout regularisation, with adding Gaussian noise, with the four strategies separately and the four strategies combined. \\
ra: remix attacks, rn: remix noise, t: transposition, t noc: transposition without time compensation}\label{tab:drum_transcript}
\end{table}

First we describe results for the traditional regularization strategies. Using dropout of probability greater than 0 does not seem to improve the training at least for snare-drum and hi-hat. Although the results are better with probabilities of 0.25  for bass-drum than without dropout, they are still below for the other instruments. moreover the results are still below the ones obtained with remixing noise and sinusoidal parts or remixing attacks except for hi-hat.

Regularization by means of adding  Gaussian noise to the input spectrograms is beneficial for all instruments, but does not achieve best results for bass-drum and snare-drum. On the other hand, adding noise is the most effective augmentation strategy for hi-hat outperforming the results with all other augmentation methods for all noise levels. The best result for hi-hat is in fact  achieved by means of adding Gaussian noise of standard deviation of 0.2. 

Concerning data augmentation with the four signal transformation strategies introduced above it becomes clear that results are different for bass-drum and snare-drum (vibrating membranes with tunable resonances) and hi-hat (no resonance). 

We first note that augmentation by means of transposition with time compensation always considerably increases recall, but at the same time decreases precision, to an extend that for the hi-hat the F-measure overall even decreases. This effect might indicate that the transposition up to 300 cents is to strong, even if perceptually the transformed sounds did still sound acceptable. The worst effect is observed for the hi-hat. It might be explained by the fact that the hi-hat is the drum instrument with the highest frequency range. So the same transposition leads to significantly increased frequency offset. It seems necessary to run instrument dependent adaptation of the transposition extents to get the best effect of the transposition for all instruments. 

On the other hand, the transposition without time compensation does not improve only the recall for the bass-drum but it improves considerably the precision too. That leads to the best results we get over all augmentation strategies. In fact, without time compensation, the transposition does not introduce artifacts. 

Augmentation by means of remixing the attack strength has a positive effect for all instruments except for hi-hat for which it degrades slightly the F-measure. 

For the snare drum, remixing noise and sinusoidal parts leads to best results. Here the F-measure increases by 0.6 points in comparison to F-measure obtained with the model trained without any augmentation. We note that for snare-drum and hi-hat, it improves precision but degrades the recall.  

It is interesting to note that augmentation with a combination of all signal transformation strategies is generally worse than the best strategy. This comforts the idea that careful adjustment of data augmentation needs to be performed taking into account the properties of the individual instruments.

\section{Conclusions}\label{sec:concl}

This article aimed to analyze the influence of adding augmented data to the training set. Moreover it compared several transformations between them as well as training with only original data and two data-independent methods which are dropout and additive Gaussian noise.

The different strategies were applied to a MIREX 2018 database for automatic drum transcription task. The first important result was the improvement of results of all experiment with at least one of the data augmentation strategies. 

In addition, for some strategies, adding augmented data did not only improve the recall as expected but also the precision. That seems to indicate that the network learns more features but chooses them more precisely too. 

However, the results with networks trained on the four strategies data combined do not reach the best results. This article shows that the data augmentation strategy and their parameters have to be chosen carefully. 

\bibliographystyle{IEEEbib}
\bibliography{biblio}

\end{document}